\documentclass[3p,times,procedia, number]{elsarticle}
\usepackage[utf8]{inputenc}
\usepackage{tabularx}
\usepackage{csquotes}

\usepackage{ecrc}

\usepackage[bookmarks=false]{hyperref}
    \hypersetup{colorlinks,
      linkcolor=blue,
      citecolor=blue,
      urlcolor=blue}

\volume{00}
\firstpage{1}
\journalname{Procedia Computer Science}

\runauth{Shukla N., Romeo A., et al.}

\jid{procs}

\usepackage{amsmath,amsfonts,amssymb}
\usepackage{bm} % Bold math

\usepackage{graphicx}
\usepackage{float}
\usepackage{caption}
\usepackage{subcaption}
\usepackage[figuresright]{rotating}

\usepackage[dvipsnames,table]{xcolor}
\usepackage{multirow}
\usepackage{stfloats}
\usepackage{placeins}

\usepackage{listings}
\usepackage{pgfplots}
\pgfplotsset{compat=1.18}

\usepackage{algorithmic}

\usepackage{url}
\usepackage{textcomp}
\usepackage{verbatim}
\usepackage{balance}
\usepackage{natbib} % for authoryear citations

\usepackage{color}
\usepackage{ulem}

\newcommand{\OG}{\texttt{OpenGadget3}}
\newcommand{\GP}{\texttt{gPLUTO}}
\newcommand{\IP}{\texttt{iPIC3D}~}

\lstnewenvironment{Ccode}{
  \lstset{
    language=C,
    emph={double},
    emphstyle={\color{black}},
    keywordstyle=[2]{\color{red}},
    keywords=[2]{pragma,acc},
    directivestyle=\color{brown},
    commentstyle=\color{blue},
    basicstyle=\ttfamily\scriptsize,
    tabsize=2,
    breaklines=true
  }
}{}

\begin{document}
\begin{frontmatter}

\dochead{ Proceedings of the Third EuroHPC user day}%%%
\title{Towards Exascale Computing for Astrophysical Simulation Leveraging the Leonardo EuroHPC System}

\author[cin]{Nitin Shukla\corref{cor1}}
\ead{n.shukla@cineca.it}
\author[cin]{Alessandro Romeo\corref{cor1}}
\ead{a.romeo@cineca.it}
\author[cin]{Caterina Caravita}
\author[cin]{Michael Redenti}
\author[it4]{Radim Vavrik}
\author[it4]{Lubomir Riha}
\author[uni]{Andrea Mignone}
\author[uni]{Marco Rossazza}
\author[uni]{Stefano Truzzi}
\author[inaf]{Luca Tornatore}
\author[inaf]{Antonio Ragagnin}
\author[inaf]{Tiago Castro}
\author[lmu]{Geray S. Karademir}
\author[lmu]{Klaus Dolag}
\author[kul]{Pranab J. Deka}
\author[kul]{Fabio Bacchini}
\author[kul]{Rostislav-Paul Wilhelm}
\author[e4]{Daniele Gregori}
\author[e4]{Elisabetta Boella}

\cortext[cor1]{Corresponding author.}

\address[cin]{CINECA, Casalecchio di Reno, Italy}
\address[it4]{IT4Innovations, VSB-TU Ostrava, Ostrava, Czech Republic}
\address[uni]{University of Turin, Turin, Italy}
\address[inaf]{INAF - Osservatorio Astronomico di Trieste, Trieste, Italy}
\address[lmu]{Universitäts-Sternwarte, Fakultät für Physik, Ludwig-Maximilians-Universität München, Munich, Germany}
\address[kul]{KU Leuven, Leuven, Belgium}
\address[e4]{E4 Computer Engineering SpA, Scandiano, Italy}

%%%%%%%%%%%%%%%%%%%%%% ABSTRACT & KEYWORDS %%%%%%%%%%%%%%%%%%%%%%

\begin{abstract}
Developing and redesigning astrophysical, cosmological, and space plasma numerical codes for existing and next-generation accelerators is critical for enabling large-scale simulations. To address these challenges, the SPACE Center of Excellence (SPACE-CoE) fosters collaboration between scientists, code developers, and high-performance computing experts to optimize applications for the exascale era. 
This paper presents our strategy and initial results on the Leonardo system at CINECA for three flagship codes, namely \GP{}, \OG{} and \texttt{iPIC3D}, using profiling tools to analyze performance on single and multiple nodes. Preliminary tests show all three codes scale efficiently, reaching 80\% scalability up to 1,024 GPUs. 
\end{abstract}


\end{frontmatter}

\vspace*{-6pt}
%%%%%%%%%%%%%%%%%%%%%% MAIN CONTENT %%%%%%%%%%%%%%%%%%%%%%

%%%%%%%%%%%%%%%%%%%%%%%%%%%%%%%%%%%%%%%%%%%%%%%%%%%%%%%%%%%%%%%%%%%%%%%%%%%%%%%%%%%
\section{Introduction}
%%%%%%%%%%%%%%%%%%%%%%%%%%%%%%%%%%%%%%%%%%%%%%%%%%%%%%%%%%%%%%%%%%%%%%%%%%%%%%%%%%%
High-performance computing (HPC) has transformed large-scale astrophysical simulations, enabling researchers to study complex phenomena such as dark matter interactions \citep{Sarkar18, Schoeffler25}, gravitational lensing \citep{Romeo17, Croft18}, magnetic reconnection \citep{Markidis12, Mattia23, Ferro24}, X-ray emissions from galaxy clusters \citep{Borgani04}, and $\gamma$-ray bursts driven by fireball beam–plasma interactions \citep{Shukla18, Boella22}. Advances in HPC hardware, particularly in GPU architectures and high-speed interconnect technologies, are driving progress towards exascale computing ($10^{18}$ calculations per second). These improvements enable significantly larger, more detailed, and faster simulations. To fully exploit this capability, scientific codes need to adopt new programming models. The SPACE-CoE project brings together scientists, developers, and tech experts to redesign astrophysical software for exascale systems. It focuses on developing efficient, portable, and scalable applications, encouraging collaboration in the European astrophysics community through shared tools, data standards, and development practices.

The SPACE-CoE \citep{spaceCF25} focuses on adapting seven widely used open-source astrophysics codes, covering magnetohydrodynamics, cosmology, general relativity, and space plasma physics, including \GP{} \citep{Rossazza25}, \OG{} \citep{Groth23}, and \IP \citep{Markidis2010}, for which CINECA co-develops GPU-centric algorithms. While these codes are already optimized for CPU parallelism, leveraging the full power of modern EuroHPC GPU-based super computers requires further optimization to enhance performance and scalability. CINECA plays a key role in code redesign, optimization, profiling, and performance analysis, as well as supporting the project on the Leonardo system. This paper is organized as follows: Section \ref{sec:workflows} gives an overview of the codes and their POP3~\cite{pop3coe2024} analysis. Section \ref{sec:GPUstrategies} covers GPU-focused development and code validation after each change. Section \ref{sec:Benchmark} outlines our benchmarking process and software engineering practices adapted for HPC. We present profiling and performance results for each application.

%%%%%%%%%%%%%%%%%%%%%%%%%%%%%%%%%%%%%%%%%%%%%%
\section{Code development workflows}
\label{sec:workflows}
%%%%%%%%%%%%%%%%%%%%%%%%%%%%%%%%%%%%%%%%%%%%%%

In this section, we describe our collaborative work with the three European research teams involved in the development activity. We also provide a brief overview of the main features of the codes and how they leverage GPU computational capabilities. The three aforementioned astrophysical codes tackle a diverse range of scientific problems, involving different algorithmic methodologies and approaches.

\subsection{Implementation of \GP{}}
\GP{} is the GPU-enabled version of \texttt{PLUTO}, a multi-algorithm framework for solving the equations governing plasma dynamics at high Mach numbers. It supports a range of physical models, including the compressible Navier-Stokes equations, ideal Magneto Hydro Dynamics (MHD), relativistic MHD (RMHD) and resistive relativistic MHD (ResRMHD). 
The new code version is implemented in \texttt{C++} and employs Godunov-type finite volume solvers for hyperbolic and parabolic MHD conservation laws in up to three spatial dimensions, on both static and mapped grids. 
Currently, \GP{} implements approximately 65\% of the modules and features available in the original \texttt{PLUTO} code. Many of the missing features of \texttt{PLUTO}, such as adaptive mesh refinement (AMR) grids, heating and cooling mechanisms, dissipation process, etc., are being ported to the GPU. The code uses \texttt{MPI} to handle multiple tasks, while GPU acceleration is handled through \texttt{OpenACC} (Rossazza et al., submitted). GPU offloading using \texttt{OpenMP} is currently in progress and will be included in a future release of the code.

\subsection{Implementation of \OG{}}
\OG{} (Dolag et al., in prep.) is an open-source version of a \texttt{Gadget2}~\citep{Gadget2} code successor. It is a N-Body \texttt{C/C++} code and deals with several astrophysical scales. Gravitational force uses Barnes \& Hut algorithm~\citep{Barnes1986} for short range interactions and particle mesh algorithm to treat large scales, performed with the use of \texttt{FFTW3} libraries \citep{FFTW05}. The code employs an advanced smoothed particle hydrodynamics (SPH) solver \citep{Beck2016}. As an alternative hydro solver, the code is also equipped with a meshless finite mass method \citep{Groth23}. Moreover, this code also includes other baryonic physics processes, such as radiative cooling, star formation, energy feedback, radiative transfer, magnetic fields, and black holes. The code uses a hybrid \texttt{MPI} + \texttt{OpenMP} parallelization scheme, exploiting Hilbert space filling curve locality~\citep{Ragagnin16}, and both inter-node and intra-node parallelism. GPU offloading is currently implemented via \texttt{OpenACC} \citep{Ragagnin2020}, while support for \texttt{OpenMP} offloading is under development.

\subsection{Implementation of \IP}
\IP \citep{Markidis2010} is a semi-implicit Particle-in-Cell (PIC) \texttt{C/C++} code developed primarily to study collisionless plasma dynamics at kinetic scales. Macro-particles, which are used to represent an ensemble of plasma particles, are evolved in a Lagrangian framework, whereas the moments (such as plasma density, current, pressure, etc...) and the self-consistent electric and magnetic fields are tracked on an Eulerian grid. The three main kernels of \IP are Particle Mover, Moment Gatherer, and Field Solver. Due the implicit nature of the underlying algorithms, unlike explicit PIC methods, insufficiently resolved scales do not result in numerical instabilities. This allows us to choose time step sizes and spatial grid sizes 10--100 times greater than those used in traditional explicit PIC codes.  \IP is an \texttt{OpenMP} + \texttt{MPI} hybrid code, and two of its modules, the particle mover and moment gatherer, are offloaded to GPU using \texttt{HIP} (for AMD GPUs) and \texttt{CUDA} (for NVIDIA GPUs).

\subsection{Leonardo HPC system and allocation of resources}
The initial activities of the SPACE-CoE project was conducted on the EuroHPC Tier-0 supercomputer Leonardo, hosted and managed by CINECA. Leonardo consists of two main partitions \citep{turisini_etal}: the DCGP partition with 1,536 dual-socket nodes, each equipped with 112-core Intel Xeon 8480+ CPUs, 512 GB RAM, and 3.84 TB NVMe SSDs; and the Booster partition with 3,456 accelerated nodes, each with a 32-core Intel Xeon 8358 CPU and four customized NVIDIA A100 GPUs, totaling 13,824 GPUs. Each Booster node is equipped with 512 GB of memory and 64 GB of HBM2e memory per GPU. The GPUs are interconnected via NVLink 3.0 and connected to the host system through PCIe Gen4. Inter-node communication uses InfiniBand HDR with four 100 Gb/s ports per node. Leonardo Dragonfly+ network topology ensures high-bandwidth, low-latency connectivity through a two-level fat tree structure with spine and leaf switches, supporting scalable performance for large-scale simulations. Leonardo offers a pre-exascale computing capability of up to 240 Pflop/s. 

Each SPACE-CoE code needs specific libraries, managed by the SPACK package manager \cite{gamblin2015spack}. These are organized into modules for different compilers and \texttt{MPI} versions. In addition to GNU compilers, DCGP uses Intel OneAPI compilers, while Booster uses NVIDIA HPC and \texttt{CUDA} compilers for GPU tasks. Computational resources on Leonardo are provided through allocations from ISCRA and EuroHPC, enabling jobs of up to 128 nodes on DCGP and up to 256 nodes on Booster. Additionally, CINECA has allowed larger runs during special \textit{pre-exascale days}.

\subsection{CPU performance assessment with \texttt{Extrae}}
This section focuses on evaluating CPU code performance prior to any GPU adaptation. As the first step, we assessed the current status of the codes using the HPC performance analysis suite based on \texttt{Extrae} \citep{extrae}. The profiling results, obtained in collaboration with IT4I, provide insights into code behavior and performance bottlenecks, forming the basis for optimization and porting to the Leonardo HPC system.

The performance analysis process is similar for most project codes. It begins with carefully selecting and setting up test cases for measurement. To get the most useful results of the analysis for the following optimization work, it is important to use a production-level test case that performs all the required computational routines ideally at the target scale in terms of computational resources, i.e. number of nodes, CPU cores, GPU accelerators, etc. At that scale, the typical simulation runs with the duration of many hours would generate an enormous amount of performance data. To decrease the size of the collected data to a manageable level, it is necessary to limit the simulations only to several units or tens of iterations, e.g. time steps, keeping the target scale.

Among the three CoE codes only \OG{} and \IP{} exhibited scalability issues with potential for improvement. In contrast, the CPU version of \GP{} (\texttt{PLUTO}) had already undergone some optimization before the start of the SPACE project. As a result, no further scalability analysis or optimization was required for \texttt{PLUTO} at the beginning of the SPACE project.

Manual navigation and identification of relevant sections of complex simulation codes through the collected performance data tend to be very time-consuming work and prone to inaccuracies. Both issues might be addressed by instrumenting the selected regions of interest (RoI), meaning inserting simple non-intrusive probes of the particular tracing tool into the source code, typically at the beginning and end of the region.

Such an instrumented code and configured test case are used to perform the actual tracing. For \OG{} and \IP{} analysis the following tool-chains were used on Leonardo DCGP: \texttt{GCC 12.2.0}, \texttt{OpenMPI 4.1.6}, \texttt{Extrae 4.0.6}, \texttt{HDF5 1.14}, \texttt{FFTW 3.3.10} (OG), \texttt{OpenBLAS 0.3.24} (OG), \texttt{GSL 2.7.1} (OG), \texttt{PETSc 3.21.1} (IP).

\subsubsection{\OG}
\begin{figure}[H]
    \centering
    \begin{subfigure}{0.39\linewidth}
        \centering
        \includegraphics[width=\linewidth]{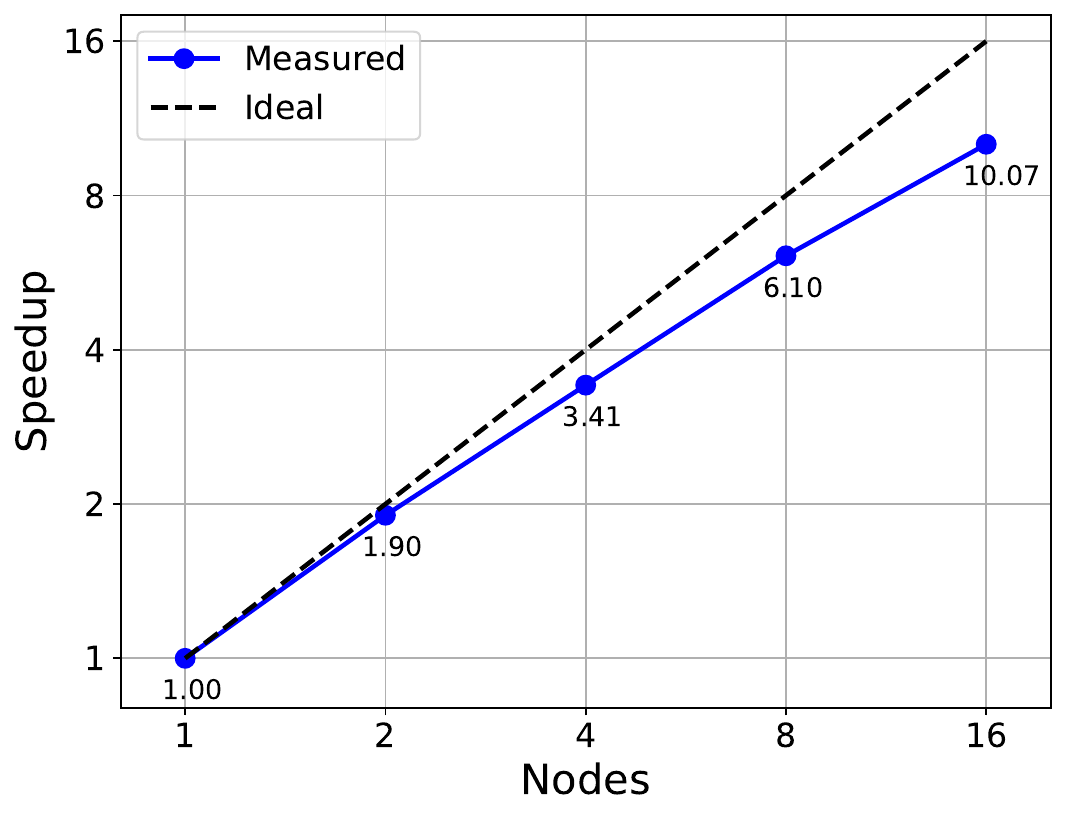}
    \end{subfigure}
    \begin{subfigure}{0.6\linewidth}
        \centering
        \includegraphics[width=\linewidth]{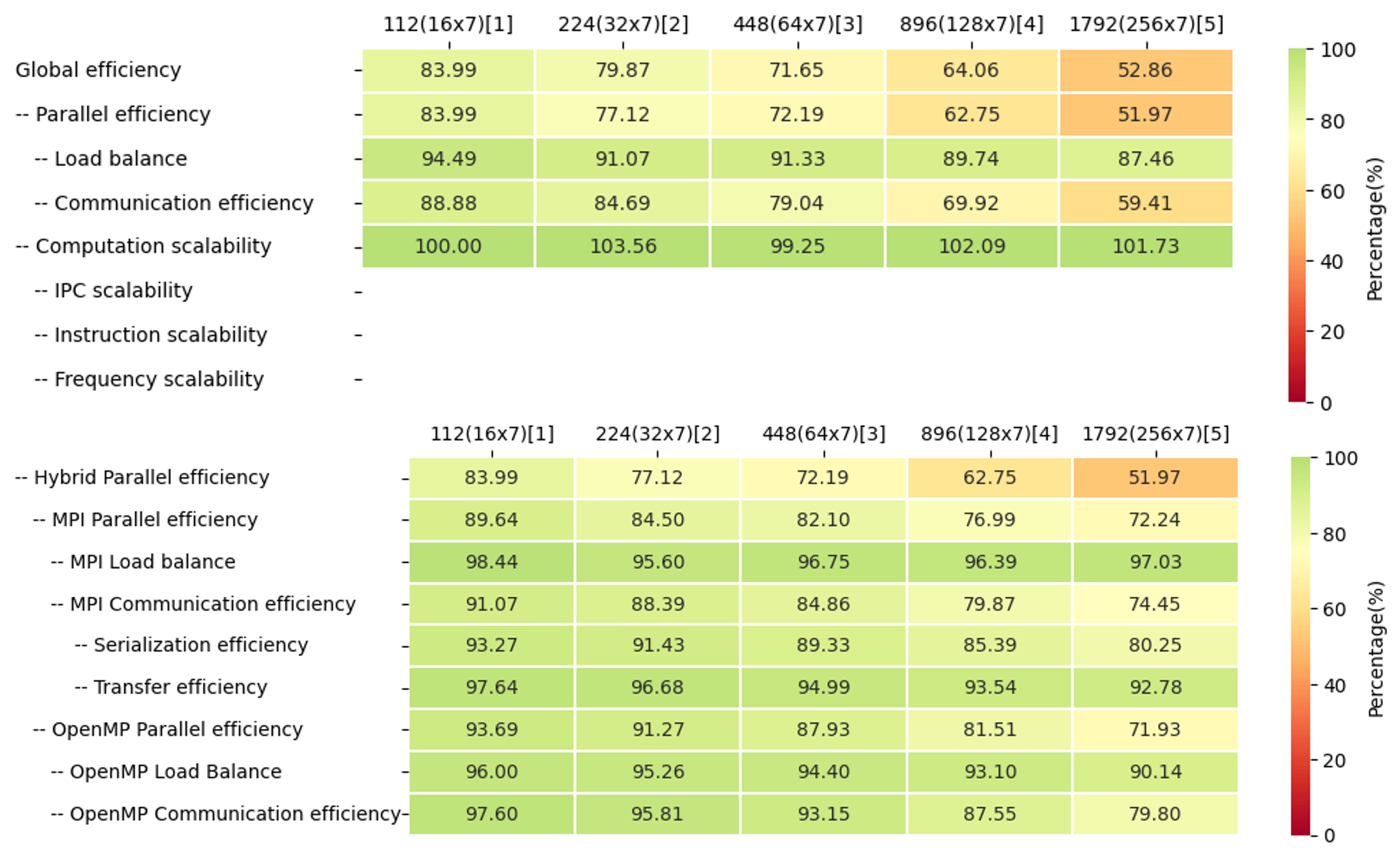}
    \end{subfigure}
    \caption{Extrae anlysis of \OG{} on Leonardo DCGP for up to 16 nodes: strong scaling (left panel) and efficiency metrics [\%] (right panel). Note that the computation scalability sub-metrics were unavailable at this time due to incompatible tool suite components.}
    \label{fig:og_pop_metrics}
\end{figure}

In the case of \OG{}, we simulated a box with a cosmological size of 30 Mpc/h and \(256^3\) particles were used. For this analysis, simulations were performed using from 1 to 16 nodes, where both gravity and hydrodynamics were enabled. The last 15 time steps were selected as RoI. Figure \ref{fig:og_pop_metrics} left panel shows that the speedup of RoI gradually diverges from the optimum, with the efficiency below 80\% on 8 nodes (896 \texttt{MPI} processes and 7 \texttt{OpenMP} threads per process). The hierarchical multiplicative model of efficiency metrics on Figure \ref{fig:og_pop_metrics} right panel identifies Serialization efficiencies (SerE) and \texttt{OpenMP} Communication efficiencies (OCE)\footnote{See \url{https://co-design.pop-coe.eu/metrics/hm/omp_communication_efficiency.html}} as the main limiting factor of the RoI scaling. The detailed explanation of the model and the metrics can be found at POP3 CoE learning materials \cite{pop3coe2024}.

Each RoI time step consists of the following set of high-level routines that were instrumented and further analyzed: Domain decomposition intensity decision (DD) subdivided in the decision criteria computation (DD1), and the actual execution (DD2), which can either perform a full domain decomposition or only transfer particles that moved out of their domain; Gravitational accelerations (GRAV); Densities (DENS); Hydro-accelerations (HYDRO); Non-standard physics (PHYS). 

The SerE describes any loss of efficiency due to dependencies between processes causing alternating processes to wait. The lowest SerE (80\%) was observed in the DD2 routine; however, due to its relatively short execution time, this routine is not the primary contributor to performance degradation in this test case. The GRAV routine, although the most time-consuming, exhibits a high SerE of 92\%, making it a low priority for SerE optimization. In contrast, the DENS routine, being the second longest and having a SerE of 86\%, emerges as the most promising candidate for optimization. A simulation assuming an ideal network with infinite bandwidth and zero latency, where messages are transferred instantaneously, showed that each process spends approximately $12\%$ of its runtime waiting in the \texttt{MPI\_Alltoall} function for dependent processes.

Regarding the OCE, which captures the synchronization and scheduling overhead induced by the \texttt{OpenMP constructs}, the lowest values 36\% and 24\% were observed in DD1 and DD2 respectively, though, being the shortest routines, they are probably not worth optimizing with the given test case. On the other hand, rather small room for improvement remains in the longest GRAV routine showing the best OCE 93\%. The best candidate for optimization is again the DENS routine with OCE 80\%, followed by HYDRO with 88\% and much shorter PHYS with 67\%. In DENS, each process spends a significant part of the total execution time in \texttt{OpenMP} runtime due to very small granularity of the parallel functions, mostly in \texttt{find\_hsml} and \texttt{compute\_unified\_gradients} functions with 24\% and 10\% in average, respectively.

\subsubsection{\IP}
\begin{figure}[H]
    \centering
    \begin{subfigure}{0.36\linewidth}
        \centering
        \includegraphics[width=\linewidth]{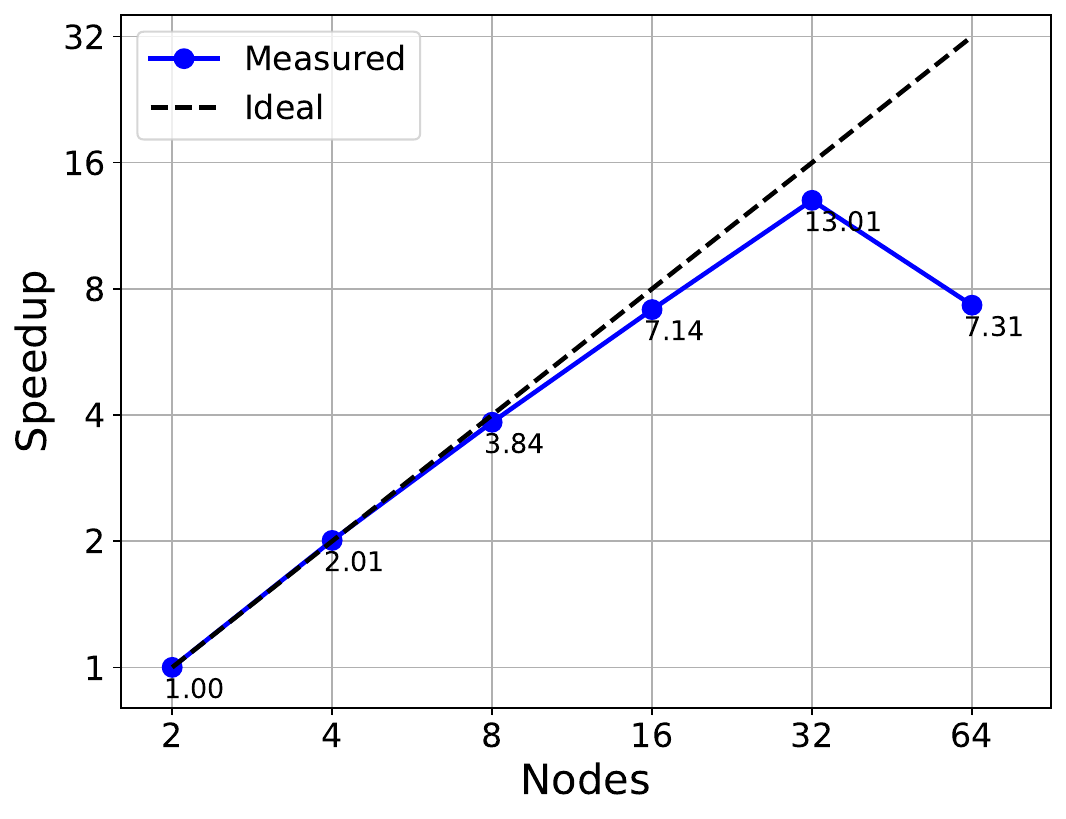}
    \end{subfigure}
    \begin{subfigure}{0.6\linewidth}
        \centering
        \includegraphics[width=\linewidth]{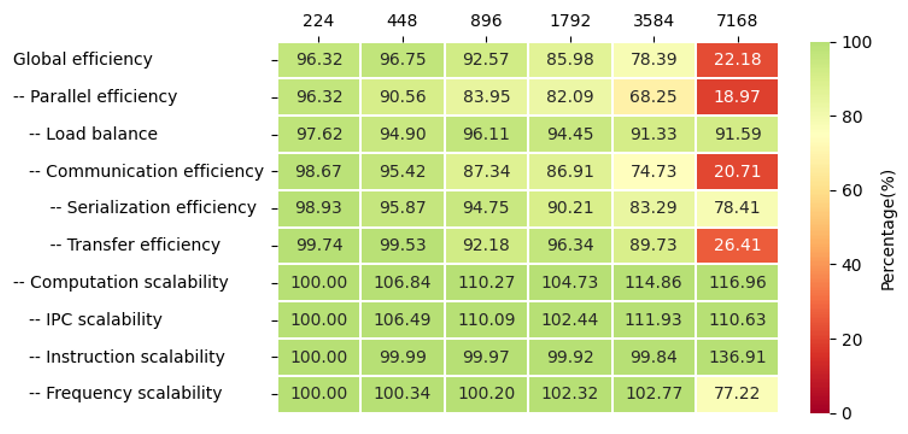}
    \end{subfigure}
    \caption{Extrae anlysis of \IP{} on Leonardo DCGP for up to 64 nodes: strong scaling (left panel) and efficiency metrics [\%] (right panel).}
    \label{fig:ip_pop_metrics}
\end{figure}

We consider the test case of a Maxwellian distribution with \(95^2\) particles per cell with 2 species on a 2D grid with a resolution of 2240$\times$1120 cells. The simulations were limited to four cycles and were conducted using between 2 and 64 nodes, with the second cycle selected as RoI. The strong scaling test in Figure \ref{fig:ip_pop_metrics} left panel reveals a clear speedup degradation on 64 nodes (7168 processes). The efficiency metrics in the right panel of Figure \ref{fig:ip_pop_metrics} point to drop in Transfer efficiency (TE) caused by an extreme growth of time in \texttt{MPI} communication and its latency.

Each cycle (RoI) is composed by the three high-level routines: CalculateField, ParticlesMover, and GatherMoments. The strongest TE deterioration at 64 nodes comes from CalculateField with 2\%, followed by GatherMoments with 41\%. That 2\% TE together with 20\% of Instruction scalability and other inefficiencies translates, in practice, into $0.2 \times$ speedup, i.e.\ $5 \times$ slowdown of the routine compared to the base run. This is caused by an excessive use of \texttt{MPI\_Barriers} bounding sequences of very small \texttt{MPI\_Sendrecv\_replace} messages with no computation in between, resulting in significant accumulated latency of the \texttt{MPI} calls. Conversely, the ParticlesMover shows very good efficiencies and almost perfect scaling. In all routines, very high sensitivity for the system preemption is also detected.

\section{Specific GPU offloading strategies}
\label{sec:GPUstrategies}
To leverage the compute capability of GPUs, all three codes adopt different numerical techniques, as each code is unique and uses distinct algorithms, each with its own set of challenges, performance bottlenecks, and numerical requirements. Additionally, these codes consist of numerous interlinked source files and are designed to simulate a wide range of physical mechanisms. As a result, optimizing or offloading to the GPU is not straightforward without focusing on a simplified, yet representative, core of the application. Therefore, we decided to extract mini-apps from the main applications, identify the relevant kernels, and analyze them to improve the overall performance of the full simulation.

Despite in the diversity numerical techniques, porting code to heterogeneous systems mainly relies on two principles: exposing parallelism and minimizing CPU-GPU data transfer. Data locality is key to GPU performance, as it minimizes memory transfers by running compute-heavy code on the GPU. Best practices include using private variables for thread independence and coalesced memory access for efficient bandwidth use. However, performance is still limited by numerical algorithms. For instance, \OG{} is memory-bound and needs new algorithms to fully utilize GPUs, while \GP{} and \IP{} have already offloaded their most compute-intensive tasks.

\subsection{Profiling activity}
We used NVIDIA Nsight tools and POP3 CoE resources to profile and analyze code performance, starting with CPU-only optimizations informed by POP3 analysis \citep{extrae, pop3coe2024}. Our strategy focused on minimizing CPU-GPU data transfers, optimizing kernel occupancy, and improving work distribution to reduce register spilling. We used profiling tools to identify performance-critical sections, analyzed memory access for coalesced patterns, and reduced data transfer overhead by overlapping communication with computation using asynchronous streams. Further, we optimized Streaming Multiprocessor (SM) occupancy and workload balance by investigating warp efficiency and control flow divergence, and reduced synchronization overhead by minimizing global atomics and barriers. This comprehensive approach guided iterative optimizations, enabling us to address bottlenecks and achieve high performance on the Leonardo Booster system \citep{SPACE_D1_2}.

\subsection{Porting of \GP{}}
The core of the \GP{} algorithm involves five main steps, which are typically repeated as many times as the number of stages employed by the Runge-Kutta solver.
These steps include: boundary exchange/calculation, mapping of the conservative vectors to primitive vectors, reconstruction (i.e. interpolation) of the cells interfaces values, solving Riemann problem to calculate the fluxes at the interfaces, computing the right hand side of the conservation law.
The divergence-free condition is controlled through the choice of one among different (mutually exclusive) algorithms (e.g. constrained transport, divergence cleaning, etc...). 
The only function that involves CPUs (or GPUs) intercommunication is the boundary exchange; this one, as well as all the other offloaded functions, was profiled and optimized for GPU parallelization: functions have been fine-tuned for coalesced memory access using a custom Array class (as shown in \citep{Rossazza25}), ensuring that independent threads access consecutive and unique memory addresses in both read and write operations. 
Managed memory is used to simplify memory management between the CPU (host) and GPU (device). 
\GP{} also handles boundary conditions and ghost cells with non-blocking \texttt{MPI} communications across domains, using asynchronous data movement across multiple GPUs. 
This setup enables fast communication with nearby tasks (8 neighbors in 2D, 26 in 3D) and fills ghost zones between processes. 
Memory is optimized by inlining frequently used functions to reduce jumps and load/store costs, while also helping the compiler access memory efficiently in GPU shared memory. 
Memory paths for multi-dimensional arrays are also determined at compile time via C++ templated functions.

\subsection{Porting of \OG}
In \OG, the N-body gravitational problem is solved using the Barnes \& Hut algorithm, which employs a hierarchical oct-tree structure to compute gravitational forces. While the benefit is a relatively low complexity of 
$O(n\log n)$ \citep{Barnes1986}, the method is memory-bound and poorly suited for GPU offloading due to irregular tree traversal, uncoalesced memory accesses, and frequent data transfers triggered by tree reconstruction at each simulation step. 
The recursive implementation of the tree build also introduces thread divergence and performance bottlenecks on GPUs, despite the Peano-Hilbert decomposition improving spatial and memory locality. Further inefficiencies arise because only a subset of particles is active at each step, requiring reordering and additional condition checks that degrade memory coalescence and warp efficiency. Previous GPU-friendly adaptations \citep{BURTSCHER201175} have not fully addressed these limitations or captured the complexity of \OG{} cosmological simulations. To address this, a new approach is under development that groups particles by spatial and memory locality (e.g., using a common center of mass), limits computations to nearby particles within small Hilbert curve segments, and uses direct summation within a fixed radius (Tornatore et al., in prep.). This reduces conditional branching, minimizes data movement, and improves warp occupancy. Additionally, non-essential data structure members were removed to reduce register spilling and enhance GPU thread performance. A more detailed explanation of this method will be provided in a forthcoming publication.

\subsection{Porting of \IP}
The scalability of \IP depends significantly on the number of particles employed, which determines the runtime of the Particle Mover (updating position and velocity of the particles at each cycle). The Moment Gatherer module interpolates particle attributes to the grid (and vice versa) to compute charge and current densities and the pressure tensor. Both the Particle Mover and Moment Gatherer tend to be compute-bound for most input configurations. This makes them highly suitable for GPUs. The Field Solver, however, uses a generalized minimal residual algorithm (GMRes) to update the electromagnetic fields. GMRes is based on matrix-free Krylov subspace projections, which can easily saturate the memory on GPUs, thereby reducing the efficiency of the overall algorithm. GMRes also requires the computation of inner products of vectors, thereby necessitating multiple internode \texttt{MPI} communications. As this module is not as compute-bound as the other two, it runs exclusively on CPUs.

Following extensive optimization of the CPU-only version and the implementation of asynchronous non-blocking communication across the \texttt{MPI} processes, the Particle Mover and Moment Gatherer modules were offloaded to GPUs, while the field solver continues to run on CPU. This introduces an overhead of data communication between the GPU and CPU at every time step. However, this is an acceptable trade-off considering the speedups obtained in the computation of the GPU-offloaded modules.

\section{Analysis and discussion of performance and scalability}
\label{sec:Benchmark}
Here, we describe our benchmarking efforts, tools, and methodology for evaluating the scalability and efficiency of these three parallel codes. A key principle in our approach is selecting problem sizes that fully utilize a single computational node, ensuring meaningful measurements of parallel performance. This strategy allows us to systematically assess how the full applications scale across multiple nodes while maintaining optimal resource utilization. 

\subsection{Automation of continuous integration and benchmarking tools}
Continuous Integration and Continuous Deployment (CI/CD) are modern practices that automate the building, testing, and deployment of code. They streamline team collaboration, reduce human error, accelerate bug fixes, and enhance overall software quality. We teamed up with IT4I to set up CI/CD workflows for our three code-bases using their GitLab Runner system \citep{SPACE_D1_3}. Regular benchmarking helps us spot performance issues and make sure our programs fully use modern hardware. It also catches slowdowns when we update code or add features, keeping performance steady over time \citep{williams-roofline}. To make benchmarking easier and work across different supercomputing systems, we use ReFrame \citep{reframe-hpc}, a Python tool for creating tests and benchmarks for HPC. ReFrame keeps system details separate from test designs, letting us write flexible tests that run smoothly on any system, speeding up both code checks and performance improvements.

\subsection{Results for \GP}
Numerical benchmarks for \GP{} have been carried out on both of Leonardo main computing partitions: DCGP and Booster. Two test cases were selected for this evaluation, the 3D Orszag-Tang vortex and the propagation of a 3D circularly polarized Alfv\'en wave. The scientific background and setup details for these tests are thoroughly discussed in \citep{Rossazza25}. The Orszag–Tang vortex involves complex shock interactions and the development of small-scale structures, making it ideal for assessing both numerical stability and performance under high-resolution conditions. On the other hand the circularly polarized Alfv\'en wave test provides an exact nonlinear solution to the ideal MHD equations and requires high accuracy and minimal numerical dissipation to preserve wave characteristics over long integration times, serving as a rigorous test of computational precision.

\begin{figure}[h]
    \centering
\includegraphics[width=0.75\linewidth]{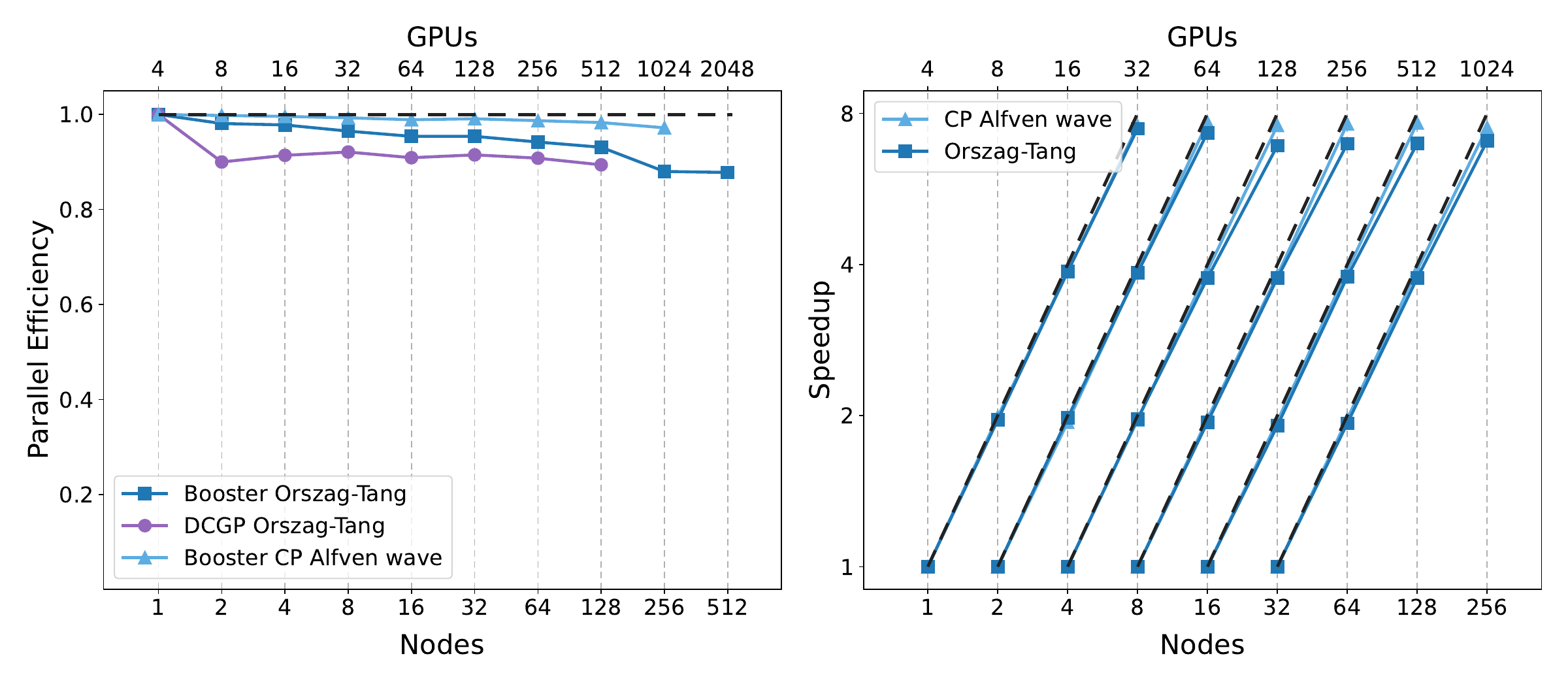}
\caption{Weak scaling tests for \GP{} on both Leonardo Booster and DCGP partitions (left panel). Strong scaling tests for \GP{} on Leonardo Booster (right panel). The black dashed lines indicate ideal scaling.}
    \label{fig:PLUTO_scal} 
\end{figure}

\textit{Weak scaling tests:}
The Orszag–Tang test simulations were conducted using a periodic 3D-Cartesian domain in double-precision arithmetic. The algorithm involved are the WENOZ reconstruction method, the HLLD Riemann solver, a third-order Runge-Kutta (RK3) time integration. To preserve the divergence-free condition of the magnetic field ($\nabla \cdot B = 0$) a constrained transport algorithm (\citep{1988_Evans}, \citep{LONDRILLO_2004}) was used. A per-node resolution of $704 \times 704 \times 352$ was selected to optimize GPU memory usage, achieving a balance between high occupancy and minimal register spilling. Figure \ref{fig:PLUTO_scal} left panel presents the parallel efficiency achieved on Leonardo Booster (up to 512 nodes) and Leonardo DCGP (up to 128 nodes) for this test, reaching approximately 88\% and 90\% efficiency, respectively. Table \ref{tab:comparison} compares absolute walltime values when running on both partitions the same weak scaling setup. Results indicate that, on average, executions on the DCGP partition are approximately an order of magnitude slower than those on the Booster partition, highlighting a significant performance gap between the two architectures.

Similar to the 3D Orszag-Tang test case, the circularly polarized Alfv\'en wave is a periodic 3D-Cartesian domain. The employed resolution is consistent with the previous Orszag-Tang test, and also in this case the computations are performed using double-precision arithmetic. Used algorithms are the same of the previous 3D Orszag-Tang test. As illustrated in the left panel of Figure \ref{fig:PLUTO_scal}, Leonardo Booster achieves an efficiency of 97\% for 256 nodes in this case.

\begin{table}[htbp!]
\vspace{1em}
\centering

\begin{minipage}[t]{0.44\linewidth}
\centering
\footnotesize
\begin{tabularx}{\linewidth}{|c|X|X|X|}
\hline
Nodes & $T_{\text{GPUs}}$ (sec) & $T_{\text{CPUs}}$ (sec) & Speedup  \\
\hline
1   & 312 & 2982 & 9.55 \\
2   & 318 & 3300 & 10.38 \\
4   & 319 & 3263 & 10.23 \\
8   & 323 & 3236 & 10.02 \\
16  & 327 & 3281 & 10.03 \\
32  & 327 & 3257 & 9.96 \\
64  & 331 & 3283 & 9.92 \\
128 & 335 & 3336 & 9.96 \\
\hline
\end{tabularx}
\caption{CPU-GPU walltime comparison of \GP{} 3D Orszag-Tang weak scaling test on Leonardo DCGP and Booster partitions.}
\label{tab:comparison}
\end{minipage}
\hspace{0.02\linewidth} 
\begin{minipage}[t]{0.44\linewidth}
\centering
\footnotesize
\begin{tabularx}{\linewidth}{|c|>{\centering\arraybackslash}p{1.2cm}|>{\centering\arraybackslash}p{1.5cm}|X|X|}
\hline
Group & Nodes  & Base Res & OT & CPA \\
\hline
1   & 1 to 8     & $832^2 \times 416$           & 0.93 & 0.96 \\
2   & 2 to 16    & $832^3$                      & 0.92 & 0.95 \\
3   & 4 to 32    & $1664 \times 832^2$          & 0.86 & 0.95 \\
4   & 8 to 64    & $1664^2 \times 832$          & 0.87 & 0.96 \\
5   & 16 to 128  & $1664^3$                     & 0.87 & 0.94 \\
6   & 32 to 256  & $3328 \times 1664^2$         & 0.88 & 0.94 \\
Mean & 1 to 256  & All res                      & 0.90 & 0.95 \\
\hline
\end{tabularx}
\caption{Strong scaling efficiencies for 3D Orszag-Tang (OT) and circularly polarized Alfvén (CPA) \GP{} simulations performed on Leonardo Booster. Each group lists minimum obtained performance relative to its base resolution and node count.}
\label{tab:strongscaling_groups}
\end{minipage}

\end{table}

\if 0
\begin{table}[htbp!]
\centering
\footnotesize
\begin{tabularx}{\linewidth}{|c|X|X|X|}
\hline
Nodes & $T_{\text{GPUs}}$ (sec) & $T_{\text{CPUs}}$ (sec) & Acceleration $(T_{\text{CPUs}}/T_{\text{GPUs}})$ \\
\hline
1   & 312 & 2982 & 9.55 \\
 \hline
2   & 318 & 3300 & 10.38 \\
 \hline
4   & 319 & 3263 & 10.23 \\
 \hline
8   & 323 & 3236 & 10.02 \\
 \hline
16  & 327 & 3281 & 10.03 \\
 \hline
32  & 327 & 3257 & 9.96 \\
 \hline
64  & 331 & 3283 & 9.92 \\
 \hline
128 & 335 & 3336 & 9.96 \\
\hline
\end{tabularx}
\caption{CPU-GPU walltime comparison of \GP{} 3D Orszag-Tang weak scaling test on Leonardo DCGP and Booster partitions.}
\label{tab:comparison}
\vspace{-1.5em}
\end{table}
\fi

\textit{Strong scaling tests:}
Strong scaling tests were performed with the same algorithm configurations of the weak scaling tests except for the technique to preserve the solenoidal condition of the magnetic field. This time the algorithm used was the divergence cleaning method (\citep{glm}, \citep{glm_1}). This allowed for choosing a grid of $(832 \times 832 \times 416)$, slightly larger than that used in the weak scaling tests. This is because the divergence cleaning algorithm consumes less device memory than the constrained transport method. It is important to note that achieving ideal GPU strong scalability becomes increasingly challenging as the number of nodes grows. This difficulty arises due to overhead associated with inter-node communication, reduced computational workload per GPU, and increased synchronization costs, all of which can limit parallel efficiency at higher node counts. To address this, we conducted six distinct strong scaling campaigns for both Orszag-Tang and circularly polarized Alf\'en problems, as detailed in Table \ref{tab:strongscaling_groups}. Within each group, the computational resolution was held constant while the number of nodes and \texttt{MPI} processes was doubled three times. This approach ensures that the problem size is sufficiently large to fully utilize GPU memory at the higher number of nodes within each group, thereby maximizing hardware occupancy and minimizing inefficiencies. Strong scaling speedup values are presented in the right panel of Figure \ref{fig:PLUTO_scal}, demonstrating nearly 89\% efficiency for the maximum number of nodes in each group for the Orszag-Tang problem, while an average efficiency of 95\% is achieved for the circularly polarized Alfv\'en problem.

\subsection{Results for \OG}
Performance tests for \OG{} were run on the Leonardo Booster and DCGP systems. Although DCGP was mainly used to test near-full-physics setups and for comparisons with the GPU versions, the main results come from the Leonardo Booster system. Since GPU support is still in progress, current results reflect only limited physics setups. Simulations model the evolution of the Universe using particles representing dark matter and gas within a 3D cosmological box. These particles interact through gravity, and gas particles also interact with hydrodynamic forces using SPH. Gravity and hydrodynamics each account for $\approx40\%$ of the runtime in a standard production run. Consequently, this setup is a good representation of the GPU performance of the entire code. High-resolution simulations require many particles and large volumes, which increase the computational demands of the short-range Barnes \& Hut part.
Simulation performance generally depends on redshift (the cosmic time). As matter is more homogeneously distributed at high redshifts, the calculation of gravitational forces via the gravity tree are faster ($\sim 10\times speedup$), compared to lower redshifts ($\sim 2\times speedup$ at $z=0$), when dense structures form, and the calculation becomes more complex and consequently more costly. So, performance varies with redshift and must be evaluated accordingly, resulting in an effective speedup of $\sim3\times$.

\begin{figure}[h]
    \centering
    \includegraphics[width=0.5\linewidth]{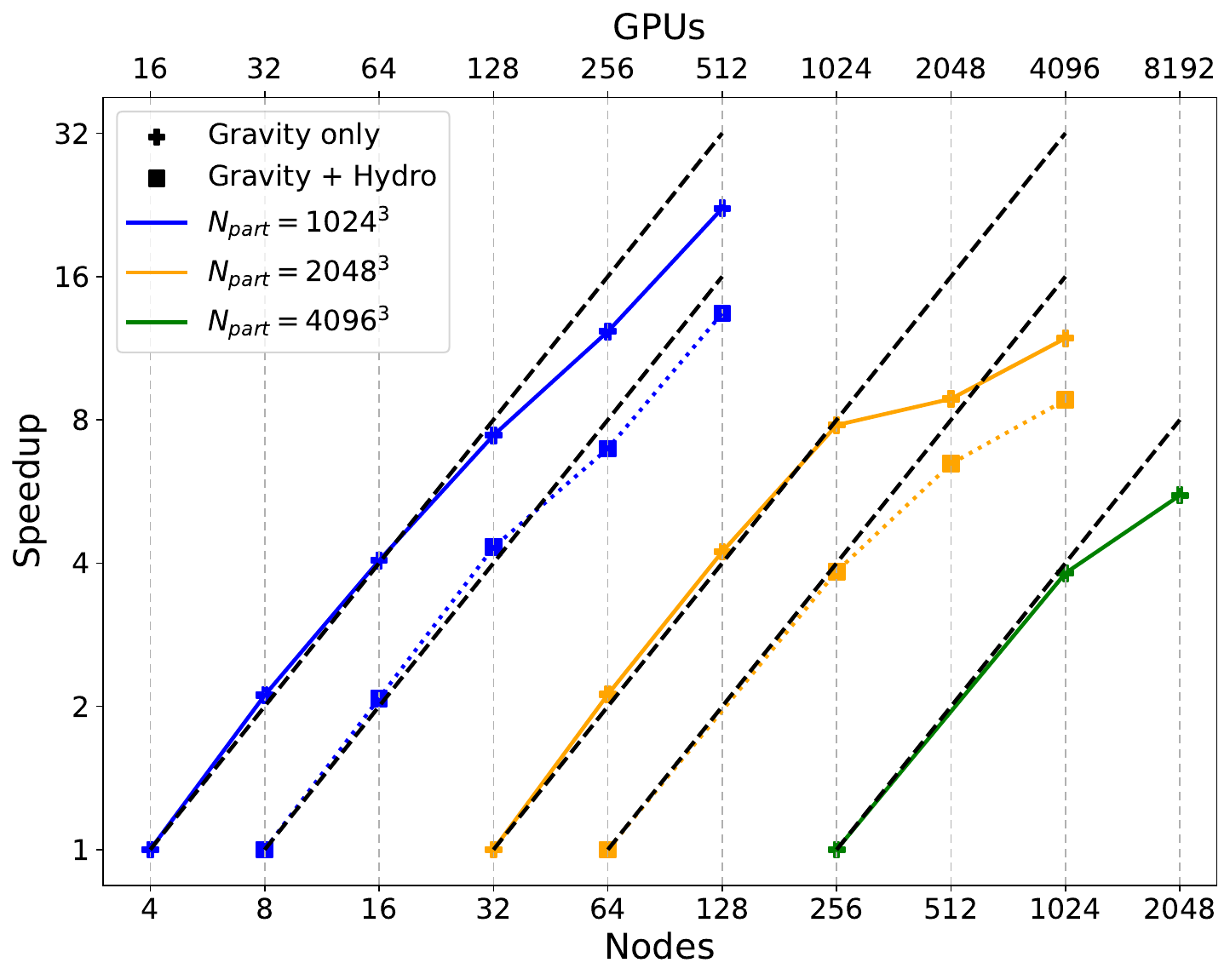}
    \caption{Strong scaling tests for \OG{} gravity-only and gravity + hydrodynamics on Leonardo Booster. The black dashed lines indicate ideal scaling.}
    \label{fig:OGstrong}
\end{figure}

We ran strong scaling tests at high redshift ($z\sim50$), using three cosmological boxes with $1024^3$, $2048^3$, $4096^3$ particles and box sizes of 120, 240, and 480 Mpc/h, respectively. This ensured the problem size was large enough per process to avoid communication issues at high node counts, which could affect scaling results. While using the same code configuration, we run these tests using initial conditions with and without gas particles. We used 4 \texttt{MPI} tasks per node (1 per GPU) and 8 \texttt{OpenMP} threads per task.

As shown in Figure \ref{fig:OGstrong}, the $1024^3$ case exhibits a smooth and consistent scaling behaviour, maintaining over $80\%$ efficiency up to a $32\times$ resource increase. The $2048^3$ case shows an efficiency drop beyond a $16\times$ scaling factor in both gravity and hydro setups, with domain decomposition issues beyond 512 nodes. While only executed in the gravity-only setup due to limited computational resources, the $4096^3$ case demonstrates speed-up of up to a $4\times$ scaling. 
In addition, a full simulation to $z = 0$ with $2048^3$ particles in gravity-only mode was performed to compare the run-times between DCGP and Booster. The right panel of Figure \ref{fig:OGstrong} shows speedup versus cosmic expansion factor $a$ (=$1/(1+z)$). Most computational time occurs at $a\gtrsim 0.2$, where efficiency drops due to the Barnes \& Hut algorithm struggling with deeper tree structures from increased clustering at lower redshifts. Smaller time steps are needed early on due to stronger gravitational accelerations, but performance gains are measured at larger $a$, yielding a 2-3 $\times$ speed-up. Despite being relatively low, this speedup is significant for simulations requiring $10^7$ or more core-hours. Since short-range gravitational force calculations cause most performance issues, this supports our approach of optimizing the tree traversals, as discussed in Section \ref{sec:GPUstrategies}.

\subsection{Results for \IP}
\begin{figure}
  \centering
    \includegraphics[width=0.5\linewidth]{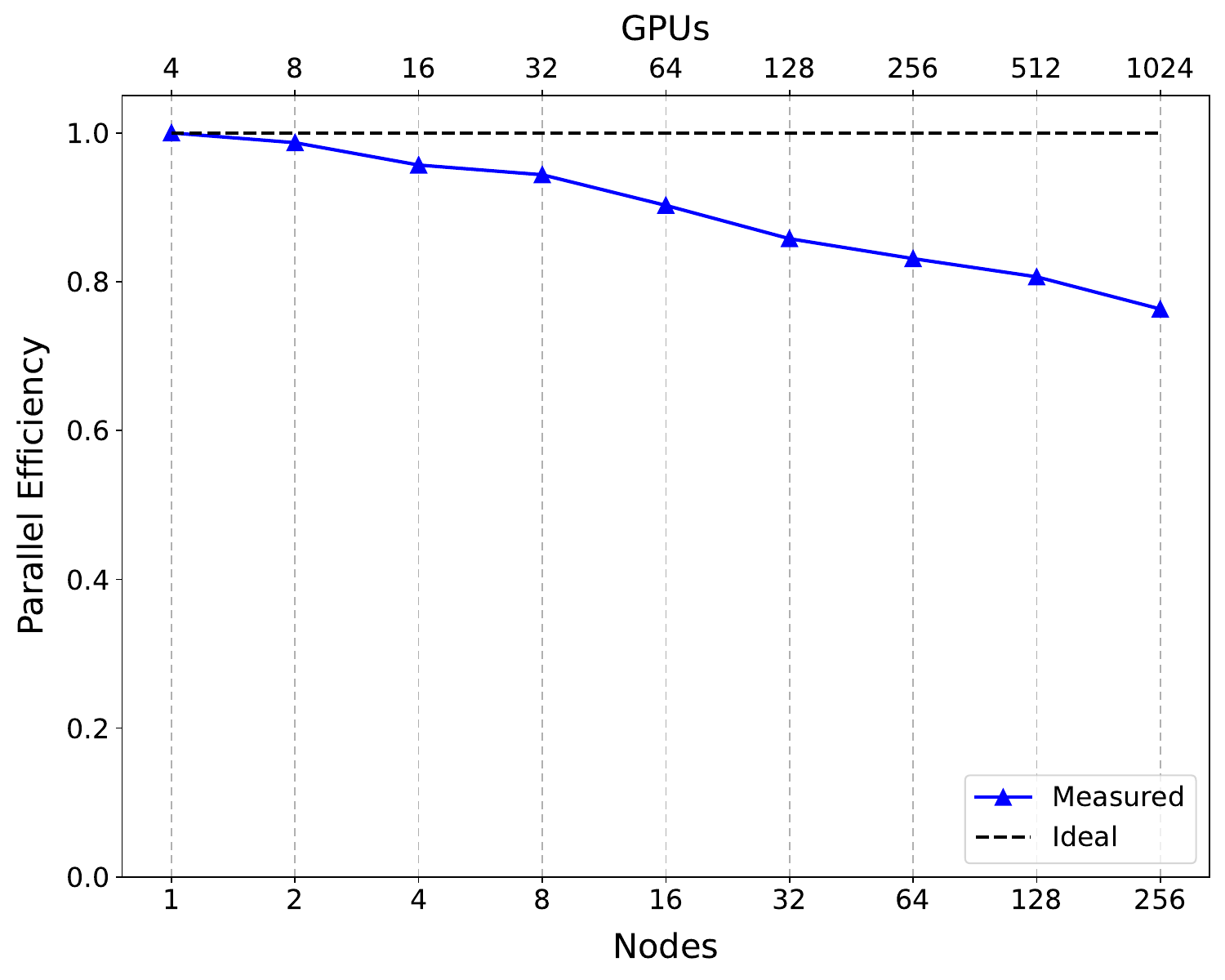}
    \caption{Weak scaling test of \texttt{iPIC3D-GPU} on Leonardo Booster. The black dashed line indicates ideal scaling.} 
    \label{fig:weak_ipic_g}
    \vspace{-1.5em} 
\end{figure}

We show a performance comparison of the CPU and GPU versions of \IP for the case of a Maxwellian distribution in 2D with $20 \times 20 \times 20$ particles per cell and 4 particle species on Leonardo DCGP and Booster, respectively (Table~\ref{tab:ipic_cg}). We observe that the Moment Gatherer obtains a speedup of 100 whereas the Particle Mover achieves a speedup of a factor of 40, which determines the overall speedup of the code. As the Field Solver runs exclusively on CPUs, no speedup is achieved for this module.
\begin{table}[htbp]
\centering
\footnotesize
\begin{tabularx}{\linewidth}{|l|X|X|c|}
\hline
\textbf{Module} & \texttt{iPIC3D-GPU} (Leonardo Booster) & \texttt{iPIC3D-CPU} (Leonardo DCGP) & \textbf{Speedup} \\
\hline
Particle Mover   & 0.542 s & 21.891 s & 40.4 \\
\hline
Moment Gatherer  & 0.123 s & 12.271 s & 99.8 \\
\hline
Field Solver     & 0.185 s & 0.183 s  & 0.98 \\
\hline
\textbf{Total}   & \textbf{0.870 s} & \textbf{35.007 s} & \textbf{40.2} \\
\hline
\end{tabularx}
\caption{Comparison of the CPU and GPU version of \IP{} for a 2D Maxwellian distribution test case with $20 \times 20 \times 20$ particles per cell and 4 total species.}
\label{tab:ipic_cg}
\end{table}
Furthermore, we illustrate the weak scaling of \texttt{iPIC3D-GPU} on Leonardo Booster, the results of which are reported in Fig.~\ref{fig:weak_ipic_g}. We consider the same Maxwellian distribution in 2D, where the number of grid cells was progressively increased from $128^2$ to $2048^2$. Each simulation modeled 4 particle species with $180 \times 180$ particles per cell per species. The code demonstrated an efficiency of $78\%$ up to 1024 GPUs, corresponding to 256 nodes on the Leonardo Booster.

%%%%%%%%%%%%%%%%%%%%%%
\section{Summary}
\label{sec:summary}
This paper presents the strategy and early achievements of the EuroHPC SPACE Center of Excellence (SPACE-CoE) in adapting three flagship astrophysical simulation codes i.e. \GP, \OG, and \IP, for exascale computing on the Leonardo supercomputer at CINECA.  Through collaborative efforts, key modules were successfully offloaded to GPUs, enabling a transition from CPU to GPU architectures using profiling and optimization techniques. Preliminary results on the EuroHPC Leonardo system show notable scalability up to 1,024 GPUs with efficiencies around 80-97\%. The study highlights \GP's near-ideal weak and strong scaling, \OG's bottlenecks with the Barnes \& Hut algorithm mitigated through restructuring, and \IP’s substantial GPU acceleration, achieving up to 100× speedups in certain modules and 78\% weak scaling efficiency by leveraging selective GPU offloading and asynchronous CPU-GPU communication. By combining performance profiling, code modularization, continuous integration, and GPU-specific kernel optimization, the project demonstrates how interdisciplinary collaboration can modernize complex simulation tools to fully exploit emerging exascale infrastructure.

%\section{Online license transfer}
%All authors are required to complete the Procedia license transfer agreement before the article can be published, which they can do online. This transfer agreement enables Elsevier to protect the copyrighted material for the authors, but does not relinquish the authors' proprietary rights. The Procedia license transfer covers the rights to reproduce and distribute the article, including reprints, photographic reproductions, microfilm or any other reproductions of similar nature and translations. Authors are responsible for obtaining from the copyright holder, the permission to reproduce any figures for which copyright exists. 

%%%%%%%%%%%%%%%%%%%%%% ACKNOWLEDGEMENTS %%%%%%%%%%%%%%%%%%%%%%

\section*{Acknowledgements}
This work was supported by the SPACE CoE, funded by the EU and several partner countries under grant No. 101093441. We also thank ISCRA and the EuroHPC Joint Undertaking for access to the computational resources on Leonardo supercomputer at CINECA (Italy).
This work was partially funded by the POP3 project (grant No. 101143931), supported by the EuroHPC Joint Undertaking, its member countries, and additional funding from the Czech Ministry of Education (ID: MC2401) and e-INFRA CZ (ID: 90254).
We also would like to acknowledge the NVIDIA experts for their invaluable assistance in porting the code to GPUs during and after dedicated hackathon events, especially Matt Bettencourt and Filippo Spiga.
%%%%%%%%%%%%%%%%%%%%%% BIBLIOGRAPHY %%%%%%%%%%%%%%%%%%%%%%

\bibliographystyle{elsarticle-num}
\bibliography{biblio}

\end{document}